\documentclass[amssymb,amsmath,pra,showpacs,twocolumn,floatfix,superscriptaddress]{revtex4}
\usepackage{graphicx}
\usepackage{lastpage}
\usepackage{amssymb}

\begin{document}

\title{Pulse shaping by coupled-cavities: Single photons and qudits}

\author{Chun-Hsu Su}
\email{chsu@ph.unimelb.edu.au}
\affiliation{Quantum Communications Victoria, School of Physics, University of Melbourne, VIC 3010, Australia}

\author{Andrew D. Greentree}
\affiliation{Quantum Communications Victoria, School of Physics, University of Melbourne, VIC 3010, Australia}

\author{William J. Munro}
\affiliation{Hewlett-Packard Laboratories, Filton Road, Stoke Gifford, Bristol BS34 8QZ, United Kingdom}
\affiliation{National Institute of Informatics, 2-1-2 Hitotsubashi, Chiyoda-ku, Tokyo 101-8430, Japan}

\author{Kae Nemoto}
\affiliation{National Institute of Informatics, 2-1-2 Hitotsubashi, Chiyoda-ku, Tokyo 101-8430, Japan}

\author{Lloyd C. L. Hollenberg}
\affiliation{Quantum Communications Victoria, School of Physics, University of Melbourne, VIC 3010, Australia}

\date{\today}

\begin{abstract}
Dynamic coupling of cavities to a quantum network is of major interest to distributed quantum information processing schemes based on cavity quantum electrodynamics. This can be achieved by actively tuning a mediating atom-cavity system. In particular, we consider the dynamic coupling between two coupled cavities, each interacting with a two-level atom, realized by tuning one of the atoms. One atom-field system can be controlled to become maximally and minimally coupled with its counterpart, allowing high fidelity excitation confinement, $Q$-switching and reversible state transport. As an application, we first show that simple tuning can lead to emission of near-Gaussian single-photon pulses that is significantly different from the usual exponential decay in a passive cavity-based system. The influences of cavity loss and atomic spontaneous emission are studied in detailed numerical simulations, showing the practicality of these schemes within the reach of current experimental solid-state technology. We then show that when the technique is employed to an extended coupled-cavity scheme involving a multi-level atom, arbitrary temporal superposition of single photons can be engineered in a deterministic way. 
\end{abstract}


\maketitle

\section{Introduction}
Cavity quantum electrodynamics provides a natural setting for distributed quantum information processing (QIP), by bringing together matter-based quantum systems for long-term memory storage, photons for fast and reliable transport of quantum information over long distances, and high finesse cavities for strong matter-field interaction. State transfer~\cite{czkm97,pellizzari97,clark03,enk97a,enk99,biswas04,hu08a}, two-qubit gates~\cite{enk97b,zheng00,guo02,serafini06,devitt07,loock08,su08a} and entanglement generation~\cite{czkm97,zheng00,guo02,serafini06,duan01,ladd06,munro08,hu08b} can all be realized through controlled coupling of spatially distant atoms. This can be achieved by connecting the cavities in which atomic qubits reside via an optical fiber~\cite{czkm97}. In other proposals, the occupation of the fiber mode is bypassed in the adiabatic limit~\cite{enk99,serafini06} or the mode is entirely avoided by coupling two atoms with a common field mode within the same cavity~\cite{enk97b,zheng00}. Alternatively, coupled-cavity lattices consisting of weakly-coupled optical cavities each containing one or more atoms, have been proposed, since each lattice site can be individually controlled and measured. Interactions in passive, coupled two atom-cavity systems have been studied for quantum interference effects~\cite{skarja99} and atomic state transfer~\cite{nohama07,ogden08}. Large-scale doped lattices have also been studied for photonic phase transitions~\cite{hartmann06,greentree06a,angelakis07}, quantum transport~\cite{bose07,makin09}, and polaritonic cluster state preparation~\cite{angelakis08}. 

An all optical coupled-cavity waveguide is also of major interest for engineering lossless guiding, slow light and enhanced nonlinearity~\cite{yariv99,xu00}. Propagation along such a waveguide can then be regulated by controlling an atom at one site~\cite{shen05,zhou08}. This type of dynamic control is also considered in the two-cavity arrangement~\cite{greentree06b, su08a} where one atom-cavity system, sandwiched between an adjacent waveguide and another cavity (e.g., qubit cavity), is manipulated by tuning its two-level atom. As a result, the coupling between the waveguide and the qubit cavity can be turned on and off on demand. Single photons can also be prepared in this way via a reversible adiabatic process and therefore it must also act as a receiver of single-photon states~\cite{su08a}. Since the control is external to the actual qubit cavity and the latter can be made optically isolated from communication channels, the qubit cavity is able to act as a separate QIP subsystem. The scheme therefore offers flexibility in quantum network and distributed QIP and indeed, has been discussed in the context of optical networks for preparing photonic 2D and 3D cluster states~\cite{stephens08,devitt08,ionicioiu09a}.

In the present paper, we examine this scheme of dynamic two-cavity coupling, with a particular focus on improving the schemes for ultra-low loss confinement and fast switching in Sec.~II. As an application, we first show in Sec.~III that simple tuning can be used to tailor the pulse shape of single photons for mode matching. As opposed the usual decay profiles from a passive cavity-based system, the single photons prepared this way have near-Gaussian shapes over a range of cavity parameters. Being the minimum uncertainty state, such pulses would be most robust against mode mismatch and therefore ideal for two-photon interference~\cite{hom87} and optical QIP~\cite{rohde05}. We note that pulse shaping has only been studied in a single atom-cavity configuration in Refs.~\cite{fattal06,fernee07,khan08,vasilev09}. In Sec.~IV, the practicality of realizing our scheme in solid-state environment (e.g., superconducting stripline resonators with mesoscopic qubits, photonic crystals and slot-waveguides with doped impurities such as diamond colour centres) is discussed. We generalize the results of Sec.~II in Sec.~V where the technique is applied to an extended scheme involving a multi-level atom, and show arbitrary temporal superposition of single photons can be engineered in a deterministic way. Such a scheme can be considered as an integrated-photonic alternative to conventional interferometric method to preparing photonic qudits for QIP proposals such as higher-dimensional quantum bus~\cite{louis08}, quantum gate~\cite{ionicioiu09b}, error filtration~\cite{gisin05} and quantum cryptography~\cite{inoue02,buttler02,takesue07}.

\begin{figure}[tb!]
\includegraphics[width=0.9\columnwidth,clip]{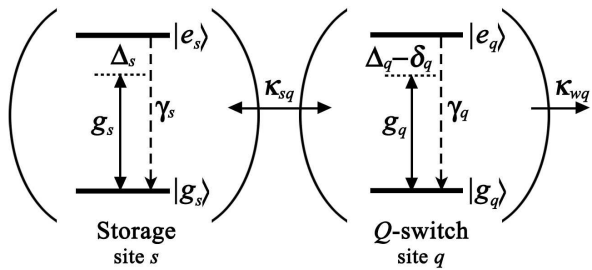}
\caption{Schematic of the two coupled atom-cavity systems for $Q$-switching and pulse shaping. $|g_\alpha\rangle$ and $|e_\alpha\rangle$ denote the ground and excited states of atom $\alpha$ ($\alpha = s, q$). Each atom with lifetime $1/\gamma_\alpha$ is coupled to the local quasi-resonant cavity mode with single-photon Rabi frequency $g_\alpha$. Atom $s$ is detuned from the local cavity mode by $\hbar\Delta_s$ whereas atom $q$ is detuned from cavity $q$ by $\hbar(\Delta_q-\delta_q)$, where $\hbar\delta_q$ is the detuning between two cavities. The cavities are evanescently coupled with rate $\kappa_{sq}$ and the right cavity is coupled to an adjacent waveguide with decay rate $\kappa_{wq}$.}
\label{fig:QSwitchScheme}
\end{figure}

\section{$Q$-switching}
An ultra-low loss (high cavity-$Q$) cavity represents an ideal environment for achieving coherent atomic manipulation at the single-quantum level with minimal dissipation. However, the fundamental time-bandwidth relation of passive low-loss cavities leads to the difficulty of out-coupling the confined light field from the cavity on demand. Furthermore, input fields must be of narrow bandwidths that limit operating speeds. This problem can be overcome using appropriate dynamic controls that modify the effective cavity-waveguide coupling, from high $Q$ that enables light confinement to low $Q$ for field in and out-coupling, with a tuning time much shorter than the photon lifetime of the cavity. This is termed $Q$-switching and here we are interested in such a control on the single-quantum level.

A schematic of our model coupled-cavity system is shown in Fig.~\ref{fig:QSwitchScheme}. It is formed by two evanescently-coupled single-modal cavities, each contains a single two-level atomic or atom-like system quasi-resonant with the local cavity mode. The atom-cavity system on the right acts as a storage (labelled with $s$) for a single quantum of excitation, whereas the left atom-cavity system ($q$) acts as a $Q$-switch to control this confinement. In turn, the switch is coupled to an external reservoir, e.g., a waveguide ($w$). Under the rotating-wave approximation, the Hamiltonian of the combined system is given by
\begin{eqnarray}
\mathcal{H} & = & \mathcal{H}_{\rm sys} + \mathcal{H}_{\rm ext}+ \mathcal{H}_{\rm int}, \\
\mathcal{H}_{\rm sys}/\hbar & = & \omega_s a_s^\dagger a_s + \nu_s |e_s\rangle\langle e_s| + g_s \Big( |e_s\rangle\langle g_s| a_s + {\rm h.c.} \Big) \nonumber \\
  & + & \omega_q a_q^\dagger a_q + \nu_q|e_q\rangle\langle e_q| +  g_q \Big( |e_q\rangle\langle g_q| a_q + {\rm h.c.} \Big) \nonumber \\
  & + & \kappa_{sq}\Big(a_{q}^\dagger a_{s} + {\rm h.c.}\Big) \label{eq:hamoriginal}, \\
\mathcal{H}_{\rm ext}/\hbar & = &  \int_{-\infty}^{\infty}{ \omega b^\dagger(\omega)b(\omega) d\omega}, \\
\mathcal{H}_{\rm int}/\hbar & = &  i\int_{-\infty}^{\infty}{\sqrt{\frac{\kappa_{wq}}{2\pi}}\left[b^\dagger(\omega)a_q - a_q^\dagger b(\omega)\right] d\omega}
\label{eq:Hint}
\end{eqnarray}
where $|g_\alpha\rangle$ denotes atomic ground state at site $\alpha = s, q$ and $|e_\alpha\rangle$ is the excited state. $a_\alpha$ and $b(\omega)$ are the bosonic annihilation operators for the excitation in cavity mode $\alpha$ and the external photon mode with frequency $\omega$, respectively. 

The physical meanings of the parameters are as follows: 
All energies are defined with respect to the resonant energy $\hbar\omega_s$ of cavity $s$, $\hbar\omega_q \equiv \hbar(\delta_q+\omega_s)$ is the resonant energy of cavity $q$, $\hbar\nu_s \equiv \hbar(\Delta_s+\omega_s)$ is the transition energy of atom $s$, and $\hbar\nu_q \equiv \hbar(\Delta_q + \omega_s$) is the corresponding energy of atom $q$. In other words, atom $q$ is detuned from cavity $q$ by $\hbar(\Delta_q - \delta_q)$. At site $\alpha$, $g_\alpha$ is the local single-photon Rabi frequency. For dipole-type atom-field interaction inside the cavity, this coupling is related to the transition dipole moment $d_\alpha$ of atom $\alpha$ and effective cavity mode volume $V_\alpha$ by $g_\alpha = d_\alpha[\omega_\alpha/(2\hbar\epsilon_0V_\alpha)]^{1/2}$. $\hbar\kappa_{sq}$ is the (photonic) couping energy between the cavity modes, and $\kappa_{wq}$ is the decay rate of the cavity mode $q$. The decay is not considered as a loss, but rather a coherent out-coupling.

Here $Q$-switching involves modifying the transmittivity of the switch -- from being reflective so that excitation is confined at site $s$, to being transmissive when this energy is allowed to propagate into the adjacent waveguide. This is achieved by tuning the transition energy of atom $q$ -- represented by $\hbar\Delta_q(t)$ as a function of time $t$. In our treatment, we assume the system is prepared in the one-quantum manifold so that the state vector of the system can be written as
\begin{eqnarray}
	|\psi(t)\rangle & = & (C_s|g_s,1_s\rangle + D_s|e_s,0_s\rangle)|g_q,0_q\rangle|{\rm vac}\rangle + \\
	& & (C_q|g_q,1_q\rangle|{\rm vac}\rangle + D_q|e_q,0_q\rangle)|g_s,0_s\rangle|{\rm vac}\rangle + \nonumber\\
	& & C_{\rm out}|g_s,0_s\rangle|g_q,0_q\rangle|\phi_w\rangle \nonumber
\end{eqnarray}
where $C_\alpha, D_\alpha$ and $C_{\rm out}$ are the probability amplitudes, the last ket denotes a vacuum or photonic state in the waveguide, and $|n_\alpha\rangle$ ($n = 0, 1$) the Fock state in cavity mode $\alpha$. From the time evolution $|\dot{\psi}(t)\rangle = (-i/\hbar)\mathcal{H}|\psi(t)\rangle$, it is straightforward to obtain the set of differential equations below,
\begin{eqnarray}
	\dot{C}_s & = & -\kappa_s/2 C_s -i\kappa_{sq} C_q - ig_s D_s, \\
	\dot{D}_s & = & (-i\Delta_s - \gamma_s/2)D_s - ig_s C_s,\\
	\dot{C}_q & = & (-i\delta -i\kappa_{sq} - \kappa_q/2 - \kappa_{wq}/2)C_q \\
	& & - ig_q D_q - \sqrt{\kappa_{wq}}f_{\rm in},\nonumber\\
	\dot{D}_q & = & (-i\Delta_q - \gamma_q/2) -ig_q C_q
\end{eqnarray}
We include the effect of atomic spontaneous emissions and incoherent cavity decays with effective terms $(-\gamma_\alpha/2)D_\alpha$ and $(-\kappa_\alpha/2)C_\alpha$ respectively, where $\gamma_\alpha$ is the transition rate of atom $\alpha$ and $\kappa_\alpha$ the transverse decay rate of cavity mode $\alpha$. The latter are related to the quality factor of the cavity, $Q_s = \omega_s/(2\kappa_s)$ and $Q_q = \omega_q/[2(\kappa_{wq}+\kappa_q)]$. $f_{\rm in}$ is the amplitude of the input pulse and is related to the output pulse $f_{\rm out}$ by the standard input-output relation~\cite{gardiner85,walls95}
\begin{equation}
	f_{\rm out} = f_{\rm in} + \sqrt{\kappa}C_q.
\end{equation}
The pulse shape of the output field is $|f_{\rm out}|^2$ and the probability of waveguide occupation is given by the integrated area of the output, $P_{\rm out} = \int_{-\infty}^{\infty} dt |f_{\rm out}(t)|^2$. 

\begin{figure}[tb!]
\includegraphics[width=0.85\columnwidth,clip]{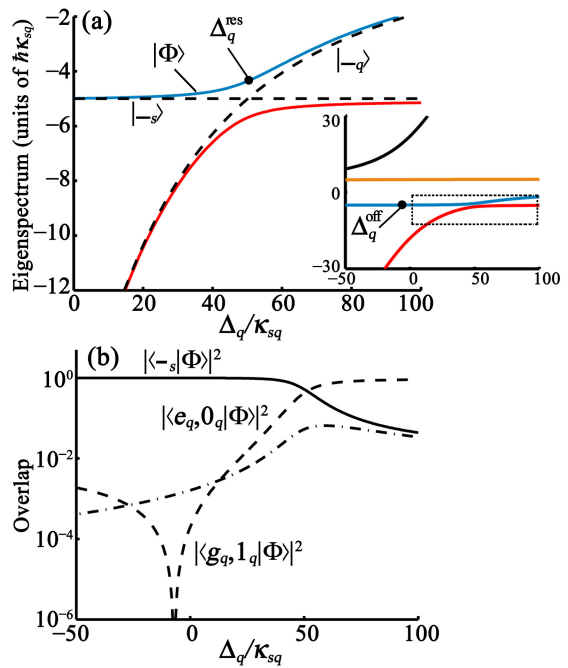}
\caption{(Color online) (a) Eigensystem of the Hamiltonian $\mathcal{H}_{\rm sys}$ (Eq.~\ref{eq:hamoriginal}) in one-quantum manifold as a function of atom-cavity detuning $\Delta_q$, for parameters $\delta_q/\kappa_{sq} = 2, g_s/\kappa_{sq} = 5$ and $g_q/\kappa_{sq} = 20$. The full view shows a closeup of the anticrossing of a particular eigenstate $|\Phi\rangle$ (blue curve) at $\Delta_q^{\rm res}$, highlighting storage-switch resonance. Dashed lines denote the energy of the dressed states of individual atom-cavity system in isolation. Inset shows the energy of all four eigenstates. (b) Overlaps showing mismatch between $|\Phi\rangle$ and state $|-_s\rangle$. Dashed, dash-dotted and solid curves corresponds to overlaps $|\langle g_q,1_q|\Phi\rangle|^2$, $|\langle e_q,0_q|\Phi\rangle|^2$, and $|\langle -_s|\Phi\rangle|^2$, respectively. The mismatch is minimized at the dip ($\Delta_q^{\rm off}$). It also shows that excitation transfers from the storage to the switch as the system evolves along $|\Phi\rangle$ from $\Delta_q^{\rm off}$ to $\Delta_q^{\rm res}$.}
\label{fig:EigOriginal}
\end{figure}

We now discuss the underlying mechanisms of $Q$-switching as follows.

\subsection{High $Q$}
A single quantum of excitation at site $s$ can be stored in the photonic $|g_s,1_s\rangle$, the atomic mode $|e_s,0_s\rangle$, or their superpositions $|\pm_s\rangle \equiv A |g_s,1_s\rangle \pm B |e_s,0_s\rangle$. In general, these dressed states can be prepared using resonant pumping and state $|g_s,1_s\rangle$ via adiabatic passage techniques~\cite{vitanov01}. Here we explictly consider the case where the system is initialized in state $|-_s\rangle$, and for simplicity, we set $\Delta_s = 0$ so that $A = -B = 1/\sqrt{2}$. 

To enable extended confinement, state $|-_s\rangle$ should be as close to one of the stationary states (i.e. energy eigenstates) of the Hamiltonian $\mathcal{H}_{\rm sys}$ as possible. For instance, we plot its eigenspectrum in Fig.~\ref{fig:EigOriginal}(a) for $\delta_q/\kappa_{sq} = 2, g_s/\kappa_{sq} = 5$ and $g_q/\kappa_{sq} = 20$ and we identify the closest eigenstate, $|\Phi\rangle$, that we use in the following discussion. Premature photonic leakage arises due to state mismatch between $|-_s\rangle$ and the prescribed eigenstate $|\Phi\rangle$. The overlap $|\langle -_s|\Phi\rangle|^2$ can be maximized in two ways. First, a very weak intercavity coupling $\kappa_{sq}$ and a large effective storage-switch detuning can be used. However, as we will see in Eq.~\ref{eq:J}, this demands a longer tuning time to out-couple this excitation. The better way is to utilize quantum interference effect. Interference can arises between two different pathways that populate state $|g_q,1_q\rangle$, from states $|-_s\rangle$ and $|e_q,0_q\rangle$ respectively. This can be thought as a spatial analogue of electromagnetically-induced transparency in a bichromatically-driven $\Lambda$ system~\cite{zhou08}. Thus, by inducing two-photon resonance between $|-_s\rangle$ and atom $q$ when
\begin{equation}
	\Delta_q^{\rm off} = \mathcal{E}_{|-_s\rangle}/\hbar,
\end{equation}
$|\Phi\rangle$ is considered as a dark state that is maximally decoupled from state $|g_q,1_q\rangle$. $\mathcal{E}_{|-_s\rangle}$ (equates $-\hbar g_s$ for $\Delta_s = 0$) is used to denote the energy of state $|-_s\rangle$. This is illustrated by the dip of the curve $|\langle |g_q,1_q|\Phi\rangle|^2$ in Fig.~\ref{fig:EigOriginal}(b). On the other hand, the other overlap is $|\langle e_q,0_q|\Phi\rangle|^2 = \kappa_{sq}^2/(2g_q^2+\kappa_{sq}^2)$, which can then be suppressed by using a large $g_q \gg \kappa_{sq}$. For $\kappa_{sq}/g_q = 10$, the leakage probability is $\mathcal{O}(10^{-3})$. Apart from this fundamental limitation, the confinement is effectively lossless when dissipation is absent.

\begin{figure}[tb!]
\includegraphics[width=0.65\columnwidth,clip]{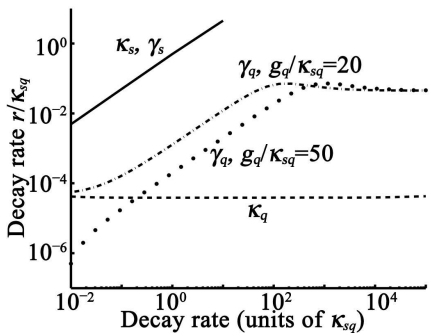}
\caption{Excitation at site $s$ dissipates at an effective rate $r$ (in units of $\kappa_{sq}$) when the condition $\Delta_q = \Delta_q^{\rm off}$ is imposed. For each curve, we only vary one of these decay rates, $\kappa_s$ (solid), $\gamma_s$ (solid), $\kappa_q$ (dashed), $\gamma_q$ (dash-dotted curve uses $g_q/\kappa_{sq} = 20$ and dotted $g_q/\kappa_{sq} = 50$), while keeping the others zero. The parameters follow Fig.~\ref{fig:EigOriginal} and $\kappa_{wq}/\kappa_{sq} = 5$ unless stated.}
\label{fig:DecohConfinement}
\end{figure}

We now examine the effect of dissipation on the confinement time when $\Delta_q = \Delta_q^{\rm off}$. Recall that there are atomic and photonic decays at site $\alpha$ with rates $\gamma_\alpha$ and $\kappa_\alpha$ respectively, we consider their contributions separately in Fig.~\ref{fig:DecohConfinement}.
For each curve, we vary one of these decay rates ($\kappa_s, \gamma_s, \kappa_q, \gamma_q$) while keeping the others zero, and calculate the effective decay rate $r$. This rate $r$ is obtained by fitting decay function ${\rm e}^{-rt}$ to the overlap $|\langle \psi(0)|\psi(t)\rangle|^2$, where $|\psi(0)\rangle = |-_s\rangle$. As expected, the confinement is most severely affected by decoherence at site $s$ with $r$ scaling with $\kappa_s$ and $\gamma_s$, and is least affected by the presence of $\kappa_q$ since state $|g_q,1_q\rangle$ is maximally decoupled. Similarly, for atomic emission ($\gamma_q$) at the switch, we compare the curves for $g_s/\kappa_{sq} = 20$ and $50$ and show that increasing $g_q$ also suppresses the decay as the storage also becomes increasing decoupled from state $|e_q,0_q\rangle$. Interestingly, we find that the decay rate peaks about $\gamma_q/\kappa_{sq} = 10^2 - 10^3$ and a larger $\gamma_q$ improves confinement.

\subsection{Low $Q$}
\begin{figure}[tb!]
\includegraphics[width=0.7\columnwidth,clip]{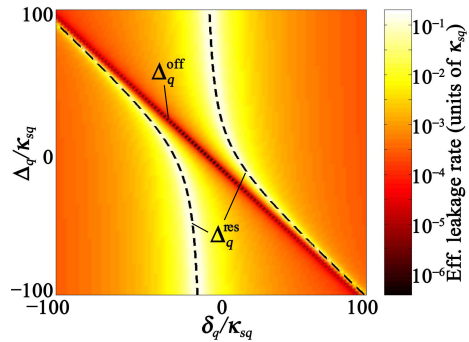}
\caption{(Color online) Light leakage rate from the storage via the switch in the passive coupled-cavity arrangement. Cavity loss is rapid at storage-switch resonance (dashed) and is minimized at $\Delta_q^{\rm off}$. We use the parameters from Fig.~\ref{fig:EigOriginal}, $\kappa_{wq}/\kappa_{sq} = 5$ and dissipation is ignored.}
\label{fig:EffLeakagePassive}
\end{figure}

When $\Delta_q \neq \Delta_q^{\rm off}$, the confined excitation is expected to leak out at a faster rate. In Fig.~\ref{fig:EffLeakagePassive}, we solve the effective leakage rate for different values of $\Delta_q$ and $\delta_q$ when the coupled-cavity system is passive. The leakage is minimized when the confinement condition is satisfied, consistent with our earlier discussion. The leakage is most rapid when the state $|-_s\rangle$ is resonant with one of the energy eigenstates of the switch, $|\pm_q\rangle$. For the setup depicted in Figs.~\ref{fig:EigOriginal} and \ref{fig:EffLeakagePassive}, this corresponds to $\mathcal{E}_{|-_s\rangle} = \mathcal{E}_{|-_q\rangle}$ in the limit of small $\kappa_{sq}$, where the detuning condition is
\begin{equation}
	\Delta_q^{\rm res} = -g_s + \frac{g_q^2}{\delta_q + g_s}.
	\label{eq:resDq}
\end{equation}
We note that, although the strength of $\kappa_{sq}$ is taken to be comparable with other parameters in our analysis, Eq.~\ref{eq:resDq} still proves to be useful as shown in Fig.~\ref{fig:EigOriginal}.

To allow controlled out-coupling, we evolve the joint storage-switch system along the eigenstate $|\Phi\rangle$ adiabatically by tuning atom $q$ from $\Delta_q^{\rm off}$ to $\Delta_q^{\rm res}$. The excitation is transferred to the switch at $\Delta_q^{\rm res}$, represented by the reduction of the overlap $|\langle -_s|\Phi\rangle|^2$ at the anticrossing in Fig.~\ref{fig:EigOriginal}(b). As we will see in the next section, this approach enables us to modify the pulse shape of the output photon. We note that the process is intrinsically reversible and therefore the switch can also couple a propagating single-photon field into the storage cavity for confinement~\cite{su08a}.

For this adiabatic process to be successful, the rate of change $\dot{\Delta}_q$ over $\Delta_q^{\rm res}$ should be slow compared to the (photon-hopping) coupling matrix element $J$,
\begin{equation}
	J \equiv |\langle -_q|(\kappa_{sq}a_q^\dagger a_s)|-_s\rangle| = \frac{\kappa_{sq}}{\sqrt{2}} \sin{\Theta}
	\label{eq:J}
\end{equation}
where $\Theta = \frac{1}{2}\arctan{[2g_q/(\Delta_q-\delta_q)]}$, which approaches zero for large $|\Delta_q-\delta_q|$. Following this, it is easy to see that decreasing cavity rate $\kappa_{sq}$ or increasing detuning $\delta_q$ implies a longer switching time is required. We can also use the standard adiabaticity criterion~\cite{shore90},
\begin{eqnarray}
	\mathcal{A} \equiv {\rm max}\left( \frac{|\langle\Phi|\dot{\mathcal{H}}_{\rm sys}|\Phi'\rangle|}{|\langle\Phi|\mathcal{H}_{\rm sys}|\Phi\rangle-\langle\Phi'|\mathcal{H}_{\rm sys}|\Phi'\rangle|^2}\right) \ll 1,
	\label{eq:A}
	\end{eqnarray}
where $|\Phi'\rangle$ is the eigenstate closest to $|\Phi\rangle$ in energy.

\section{Pulse shaping}
\begin{figure}[tb!]
\includegraphics[width=1\columnwidth,clip]{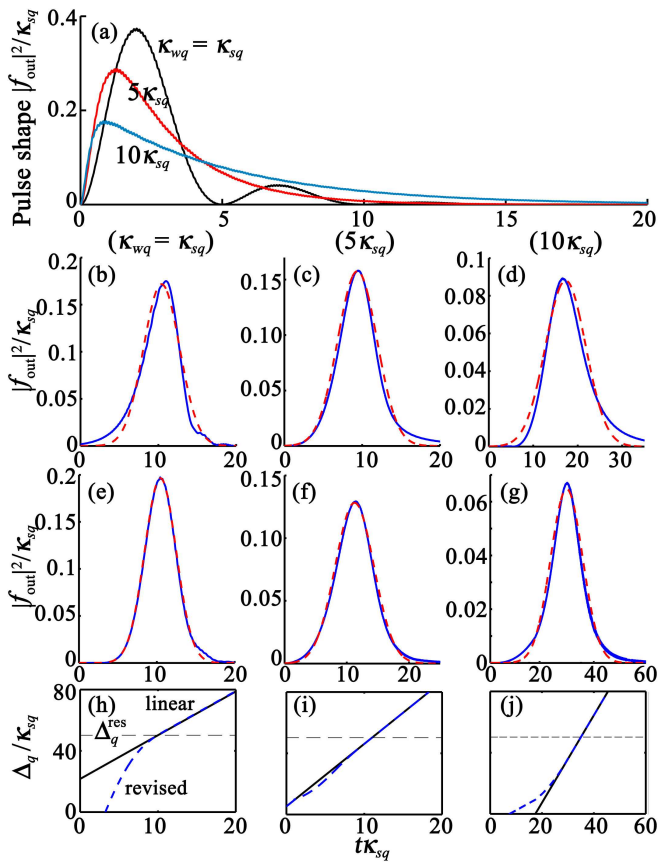}
\caption{(Color online) (a) Pulse shapes of output field for $\Delta_q =\Delta_q^{\rm res}$ when active switching is absent. Different values of $\kappa_{wq}$ are used. We then apply active switching -- varying $\Delta_q(t)$ with time $t$ from $\Delta_q^{\rm off}$ to $\Delta_q^{\rm res}$ that modifies the pulse shapes (solid line). For different values of $\kappa_{wq}$, linear sweeps are used in (b)--(d) whereas revised sweeps in (e)--(g). The corresponding sweep profiles are depicted in (h)--(j). The parameters follow Fig.~\ref{fig:EigOriginal}, dissipation is ignored and the system is initialized in state $|-_s\rangle$. Dashed lines are Gaussian fits with unit areas.}
\label{fig:PulseShaping}
\end{figure}

Custom shaping of single photons is of importance for generating suitable photons for numerous QIP applications. In particular, Gaussian pulses have shown to be optimally tolerant against mode mismatch in interference-based optical QIP schemes~\cite{rohde05}; also optimal mode-matching leads to minimal dispersion pulses, useful for long-range communication. However, photons leaked from any passive cavity-based system have non-Gaussian pulse profiles that are not robust against noise. To illustrate, we plot in Fig.~\ref{fig:PulseShaping}(a) the profile of the outgoing photon from the coupled-cavity system on resonance ($\Delta_q = \Delta_q^{\rm res}$) for different values of cavity decay rate $\kappa_{wq}$. For small $\kappa_{wq}$, the output field shows oscillations indicative of the strongly-coupled system. Critical damping occurs at $\kappa_{wq}/\kappa_{sq} \approx 2$ in this case. As $\kappa_{wq}$ increases, we observe the output becomes increasingly asymmetric. To overcome this and produce near-Gaussian pulse shapes, we are interested in tuning atom $q$ along appropriate trajectories while the confined excitation at site $s$ is out-coupled. 

Before proceeding, we note that dynamic controls for pulse shaping have only been discussed in a single atom-cavity arrangement~\cite{fattal06,fernee07,khan08,vasilev09}. In contrast, our two-cavity approach separates the two functional elements, interaction zone (site $s$) and coupling control ($Q$-switch) to allow for greater flexibility and locality of control in quantum networks. 

We first ignore decoherence and consider a linear sweep $\Delta_q(t)$ that varies from $\Delta_q^{\rm off}$ to $(\Delta_q^{\rm res} + \Delta_q^{\rm off})/2$ for each $\kappa_{wq}$ in Figs.~\ref{fig:PulseShaping}(b)--(d), respectively. The switching times $T$ between $10/\kappa_{sq}$ and $40/\kappa_{sq}$ are chosen so that the adiabaticity condition is satisfied ($\mathcal{A}\sim10^{-2}$, $J = 0.2$) and the profiles are near-Gaussian in each case. Notably, the tuning time is much shorter than the the effective lifetime at site $s$. In general, we find that larger decay rates require longer $T$. When the sweep time is fast compared to the effective leakage rate of the passive case (Fig~\ref{fig:EffLeakagePassive}), the temporal widths of the output are of the same order $O(T)$.

\begin{figure}[tb!]
\includegraphics[width=0.7\columnwidth,clip]{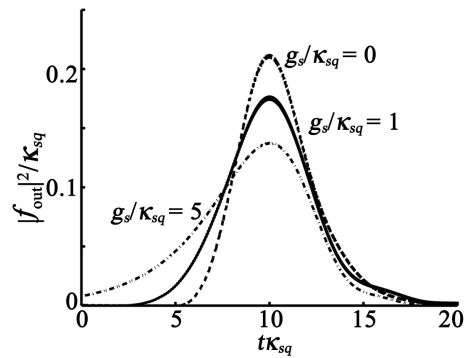}
\caption{Pulse shapes of the output photon using linear sweeps for different values of $g_s/\kappa_{sq} =$ 0 (dashed curve), 1 (solid) and 5 (dash-dotted). When atom $s$ is omitted from the system ($g_s=0$), we initialized the system in state $|1_s\rangle$. In all cases, $\delta_q/\kappa_{sq} = 2, g_q/\kappa_{sq} = 20$, $\kappa_{wq}/\kappa_{sq} = 1$, and dissipation is ignored.}
\label{fig:DifferentShapes}
\end{figure}

The premature leakage probability is $10^{-3}$ and the successful out-coupling probability is $P_{\rm out} > 0.99$. The mismatch between the output pulse $|f_{\rm out}|^2$ and an unit Gaussian fit $|g(t)|^2$ can be measured by the overlap integral $\xi \equiv 1 - \int |f_{\rm out}(t)||g(t)|dt = 0.014, 0.033$ and $0.040$ for $\kappa_{wq}/\kappa_{sq} = 1, 5$ and 10, respectively. To further improve these profiles, a more sophisticated tuning is required. Since the rate at which the switch (specifically state $|g_q,1_q\rangle$) is populated is related to the output profile, we can modify the output by changing the rate $\dot{\Delta}_q$. Using a map between a trial linear sweep and its corresponding output envelope, we reconstruct a different sweep for each case and produce results in Fig.~\ref{fig:PulseShaping}(e)--(g), with improved $\xi = 0.003, 0.01$ and 0.02, respectively. The required length of the sweep is not significantly increased and out-coupling probability remains $> 0.99$. We also examine the effect of smaller $g_s$ in Fig.~\ref{fig:DifferentShapes} and show small improvement to switching time and pulse width. In the limit of $g_s = 0$ where atom $s$ is decoupled from the system, a pulse width of $\sim5/\kappa_{sq}$ is possible because the matrix element $J$ (Eq.~\ref{eq:J}) is enhanced by a factor of $\sqrt{2}$. This is equivalent to the case where the excitation is initialized in the photonic mode $|g_s,1_s\rangle$.

\begin{figure}[tb!]
\includegraphics[width=0.65\columnwidth,clip]{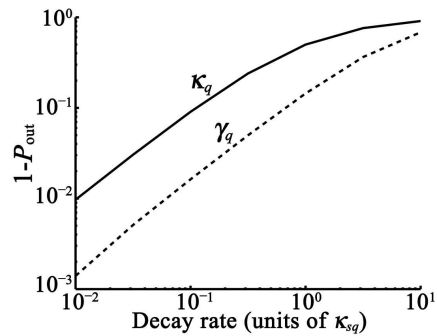}
\caption{$Q$-switching in the presence of dissipation at the switch. For each curve, we vary one of these decay rates, $\kappa_q$ (solid) and $\gamma_q$ (dashed), while keeping the other zero, and calculate the outcoupling probability $P_{\rm out}$. The parameters follow Fig.~\ref{fig:EigOriginal}(b).}
\label{fig:DecohSwitching}
\end{figure}

During $Q$-switching and pulse shaping, dissipation at the switch become important. Focusing on the loss due to spontaneous emission of atom $q$ and decay of cavity field at the switch, we plot in Fig~\ref{fig:DecohSwitching} $P_{\rm out}$ versus the decay rates $\gamma_q, \kappa_q$. The scheme is more susceptible to photon loss than spontaneous emission and degrades as the decoherence rates $\kappa_q/\kappa_{sq} > 10^{-2}$ and $\gamma_q > 0.1\kappa_{sq}$. Finally, we note that a longer sweep would not significantly reduce $P_{\rm out}$ when the sweep is much slower than the effective leakage rate $\sim 0.1\kappa_{sq}$ of the passive case (Fig~\ref{fig:EffLeakagePassive}). However, the excitation should be switched out at faster rates to avoid dissipation at site $s$.

\section{Solid-state Implementations}
Our scheme provides a mechanism for controlled coupling of atom-photon qubits in distributed QIP. In this section, we consider three possible implementations in solid-state environment and estimate their performance. Akin to many quantum information hardware proposals, one requirement for suppressing loss ($\propto \kappa_{sq}^2/g_q^2$) is strong atom-cavity coupling achieved with small cavity mode volumes. As our scheme is most suited for optical frequencies where active control of cavity-$Q$ is difficult, we first turn to an implementation in photonic-band-gap (PBG) lattice and slot-waveguide structures with diamond colour centres. We then consider superconducting striplines in the microwave region.

A defect in in PBG lattice~\cite{noda07} constitutes a wavelength-sized, high-$Q$ cavity. The proposed coupled-cavity arrangement can be formed by placing such defects in close proximity~\cite{notomi08}, and doping substitutional impurities~\cite{fu05} or quantum dots~\cite{hennessy07} as the suitable two-state atomic systems. One promising platform for solid-state quantum optics is diamond. In diamond, there are numerous colour centres with well-defined energy levels. In particular, the negatively-charged nitrogen-vacancy (NV) centre has been extensively studied for quantum electro-optical applications~\cite{greentree08}. Its zero-phonon line transition ($\omega_s, \omega_q \sim 2.95$~PHz, wavelength $\lambda = 637$~nm), between excited spin triplet state ($^3E$) to the $m=0$ sublevel of the triplet ground state ($^3A$)~\cite{manson06}, has a dipole moment $d \sim 10^{-30}$~Cm. In principle, this enables $g_s, g_q \sim 10$~GHz coupling inside a $(\lambda/2)^3$ diamond-based PBG cavity~\cite{su08b,su09} so that we estimate inter-cavity coupling $\kappa_{sq}$ of 1~GHz. With a long lifetime of 11.6~ns, $\gamma_s, \gamma_q = 86$~MHz. The transition energy of NV can be tuned by an external control field through dc Stark effect. Stark tuning of isolated centres has been demonstrated~\cite{tamarat06} and the tuning range from the earlier work of Redman \textit{et al.} of order 1~THz~\cite{redman92} is more than enough for the proposed scheme. Consequently, we estimate 10~ns operation time with $Q = 10^6$ for confinement and out-coupling probabilities $\sim 0.9$. Indeed, diamond-based PBG cavity designs in this range are available~\cite{tom06,bayn07,kreuzer08,mccutcheon08,bayn08,tom09,barclay09} and a modest quality factor of 585 has also been realized~\cite{wang07}. To further improve the fidelity to 0.99, one requires either a higher $Q$ to reduce transverse decay or a larger $g_q$ with smaller cavity mode volumes to reduce switching time.

Slot-waveguide geometries~\cite{almeida04,robinson05}, combined with mirrors, PBG or distributed Bragg (DBRs) reflectors in a Fabry-Perot arrangement~\cite{barrios05}, have also shown to be compatible with the NV and are promising for achieving ultra-small confinement~\cite{hiscocks09}. A DBR-based waveguide cavity with $Q = 27000$ has been fabricated in silicon~\cite{prussner07}, and that using PBG structures reported $Q = 58000$~\cite{velha07}. The possible subwavelength confinement predicts $g_s, g_q \sim 100$~GHz~\cite{hiscocks09} and nanosecond switching time, and such implementation only requires $Q = 10^5$ for confinement and out-coupling probabilities $>0.99$.

In the microwave regime, we can exploit strong coupling and low loss achieved in superconducting striplines. A segmented superconducting strip-line forms coupled coplanar resonators and waveguides separated by capacitive open gaps~\cite{blais04}. The capacitive atom-photon coupling at each site is implemented with a Cooper-pair box, acting as a two-state artificial atom in the charge regime~\cite{makhlin01}. Quantum superposition of its charge states can be achieved via tuning of an applied gate voltage. Its long coherence time ($1/\gamma_q, 1/\gamma_c \sim 0.1~\mu$s), large dipole moment of $10^{-25}$~Cm combined to make it well-suited to our purpose. Furthermore, the transition energy ($\sim10$~GHz) between its two levels can be tuned dynamically by changing the magnetic flux through the Josephson junction loop of the box. Since these superconducting qubits are coupled to strip-line cavities with a typical strength of $0.1$~GHz, we predict the time scale for switching with $\kappa_{sq}, \kappa_{wq} \sim 10$~MHz to be $\sim 1\mu s$ and the mode mismatch $\xi = O(10^{-3})$. To achieve confinement and out-coupling probabilities $\sim 0.9$, modest $Q = 10^4$ should suffice. This level of coupling and photon lifetimes have both been demonstrated~\cite{schuster07,majer07}. However, for enhanced probabilities $> 0.99$, photonic loss must be further suppressed with $Q = 10^5$.

\section{Preparing single photons in temporal superposition}
An integrated-photonic method to preparing photonic qudits would be a useful resource for QIP applications. Here we generalize the results in Sec. II by examining dynamic coupling between two atom-cavity systems where one of which interacts with a multi-level atom. As another application, we show that, by tuning this atom, arbitrary temporal superposition of single photons can be engineered with high fidelity. When operated in reverse, such temporally encoded qudits can be transferred into the cavity for storage or manipulation. In particular, we consider the following extensions: the multi-level atom replaces two-state atom $s$ and the atomic tuning is now performed on atom $s$. The switch (system $q$) now acts as a passive frequency filter.

\begin{figure}[tb!]
\includegraphics[width=0.9\columnwidth,clip]{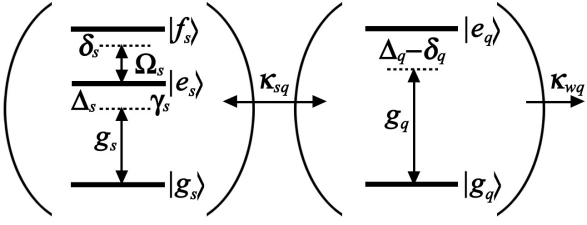}
\caption{Schematic of the extended two coupled atom-cavity systems for preparing two-time superposition states. The extensions to the previous schematic in Fig.~\ref{fig:QSwitchScheme}: atom $s$ is a three-level atom with states $|g_s\rangle, |e_s\rangle$ and $|f_s\rangle$ and the $|e_s\rangle\leftrightarrow |f_s\rangle$ transition is driven by a classical field with rate $\Omega_s$ and detuning energy $\hbar\delta_s$. As before, the $|g_s\rangle\leftrightarrow |e_s\rangle$ couples to the cavity mode with rate $g_s$ and detuning $\hbar\Delta_s$, and the cavities are evanescently coupled.}
\label{fig:SchematicThreeLevel}
\end{figure}

\begin{figure}[tb!]
\includegraphics[width=0.88\columnwidth,clip]{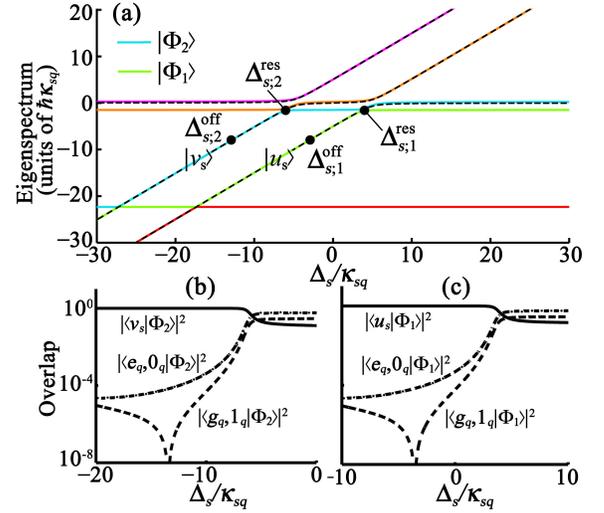}
\caption{(Color online) Eigensystem of the extended Hamiltonian $\mathcal{H}_{\rm sys}$ (Eq.~\ref{eq:ham3la}) with $\Xi$-type, three-level atom $s$ in one-quantum manifold, for $\delta_s = 0, \delta_q/\kappa_{sq}=-15, \Delta_q/\kappa_{sq} = -8.6, g_s/\kappa_{sq}=1, \Omega_s/\kappa_{sq} = 4.94$ and $g_q/\kappa_{sq}=10$. (a) Solid curves are the energies of its eigenstates, and dashed curves are the corresponding eigenenergy of system $s$ in isolation. The states $|\Phi_1\rangle, |\Phi_2\rangle$ of the joint system are approximately given by the eigenstates $|u_s\rangle, |v_s\rangle$ of system $s$ at points where their overlaps are maximum (i.e., the dips $\Delta_{s;1}^{\rm off}, \Delta_{s;2}^{\rm off}$ in (c) and (b) respectively). These overlaps decrease when the eigenstates come into resonance with the state $|+_q\rangle$ at the anticrossings ($\Delta_{s;1}^{\rm res}, \Delta_{s;2}^{\rm res}$).}
\label{fig:EigThreeLevel}
\end{figure}

\noindent \textit{Two-time superposition}:-- We begin by using a ladder-type, three-level atom with states $|g_s\rangle, |e_s\rangle$ and $|f_s\rangle$ at site $s$. The $|g_s\rangle \leftrightarrow |e_s\rangle$ transition is coupled to cavity mode $s$ with single-photon Rabi frequency $g_s$, and the $|e_s\rangle\leftrightarrow|f_s\rangle$ transition is coupled via a classical quasi-resonant field with strength $\Omega_s$ and detuning energy $\hbar\delta_s$. With the two atom-cavity systems evanescently coupled, the schematic is shown in Fig.~\ref{fig:SchematicThreeLevel}. Splitting the system Hamiltonian of Eq.~\ref{eq:hamoriginal} into components, $\mathcal{H}_{\rm sys} = \mathcal{H}_s + \mathcal{H}_q + \mathcal{P}$ where $\mathcal{H}_\alpha$ is the local Hamiltonian at site $\alpha$ and $\mathcal{P} = \hbar\kappa_{sq}(a_q^\dagger a_s + {\rm h.c.})$ is the intercavity coupling, $\mathcal{H}_s$ is now
\begin{eqnarray}
\mathcal{H}_s/\hbar & = & \Delta_s|e_s\rangle\langle e_s| + (\Delta_s+\delta_s)|f_s\rangle\langle f_s| \label{eq:ham3la}\\
&  + & g_s |e_s\rangle\langle g_s| a_s + \Omega_s|f_s\rangle\langle e_s| + {\rm h.c.}  \nonumber 
\end{eqnarray}
where the physical meanings of the other parameters are explained following Eq.~\ref{eq:Hint}, and $\Delta_s(t)$ represents the tuning of atom $s$ for controlling the effective coupling between system $s$ and $q$. For convenience, we set $\delta_s = 0$. It is instructive to see the eigenspectrum of the system Hamiltonian $\mathcal{H}_{\rm sys}$ in Fig.~\ref{fig:EigThreeLevel}. In this case, we are interested in two particular (labeled) eigenstates $|\Phi_1\rangle$ and $|\Phi_2\rangle$, which are close to two local eigenstates ($|u_s\rangle$ and $|v_s\rangle$ respectively) of system $s$ in isolation (i.e. $\mathcal{H}_s$). The analytical solutions to the doubly-dressed system $s$ can be found in Ref.~\cite{echaniz01}. 

Both states, $|\Phi_i\rangle$, anticross with other eigenstate, and by analogy, these anticrossings corresponds to resonance between system $s$ and one of the eigenstates at site $q$ (in this case, $|+_q\rangle$). In the limit of small $\kappa_{sq}$, these occur at values of $\Delta_s$ that satisfy
\begin{eqnarray}
	\mathcal{E}_{|u_s\rangle} \equiv -\frac{\hbar}{3} \left[-2\Delta_s + 2p~{\rm cos}\left(\frac{\theta}{3}+\frac{\pi}{3}\right)\right] & =&  \mathcal{E}_{|+_q\rangle},	\\
	\mathcal{E}_{|v_s\rangle} \equiv -\frac{\hbar}{3} \left[-2\Delta_s + 2p~{\rm cos}\left(\frac{\theta}{3}-\frac{\pi}{3}\right)\right] & =&  \mathcal{E}_{|+_q\rangle}
	\label{eq:doublydressed2}
\end{eqnarray}
where $\mathcal{E}_{|u_s\rangle}$, $\mathcal{E}_{|v_s\rangle}$ are the energy of states $|u_s\rangle$ and $|v_s\rangle$, $p^2 = \Delta_s^2 + 2(g_s^2+\Omega_s^2)$ and $\cos{\theta} = -(\Delta_s/p^3)[\Delta_s^2+9(g_s^2/2 - \Omega_s^2)]$. We label these resonance points $\Delta_{s;1}^{\rm res}$ and $\Delta_{s;2}^{\rm res}$. Therefore, when the system is prepared in state $|u_s\rangle$ ($|v_s\rangle$), and atom $s$ is tuned over $\Delta_{s;1}^{\rm res}$ ($\Delta_{s;2}^{\rm res}$), the confined excitation in the dressed mode is allowed to leak out via the switch, representing the \textit{low-$Q$} regime similar to the scheme previously discussed in Sec.~II. Depending on the eigenstate ($|u_s\rangle$ or $|v_s\rangle$) in which the system is initialized, a \textit{high-$Q$} regime is retrieved by inducing two-photon resonances between the eigenstate and atom $q$, as shown in the inset of Fig.~\ref{fig:EigThreeLevel}. The values of $\Delta_s = \Delta_{s;1}^{\rm off}, \Delta_{s;2}^{\rm off}$ that maximize the overlaps $|\langle u_s|\Phi_1\rangle|^2$ and $|\langle v_s|\Phi_2\rangle|^2$, satisfy the respective conditions,
\begin{equation}
	\mathcal{E}_{|u_s\rangle} = \hbar\Delta_q,~~ \mathcal{E}_{|v_s\rangle} = \hbar\Delta_q.
\end{equation}

We turn next to describe the steps for generating single-photon superposition states. Given the above choice of parameters, we initialize the system in a normalized superposition of the eigenstates, $A|u_s\rangle + B|v_s\rangle$ for $\Delta_s = \Delta_{s;2}^{\rm off}$. This can be set up, for example, using a laser pulse sequence engineered with the GRAPE algorithm~\cite{khaneja05} or by reversing the out-coupling process. This choice of $\Delta_s$ ensures confinement in mode $|v_s\rangle$. However, because the optimal confinement condition for mode $|u_s\rangle$ cannot be simultaneously imposed, we use the standard storage-switch off-resonance to effect suitable confinement in this mode by ensuring $\Delta_{s;2}^{\rm off}$ is far from resonance ($\Delta_{s;1}^{\rm res}$). Then at later times $t_B$ and $t_A$ ($0< t_B < t_A$), the resonances between state $|v_s\rangle$ with $|+_q\rangle$ and state $|u_s\rangle$ with $|+_q\rangle$, are induced by tuning atom $s$ from $\Delta_{s;2}^{\rm off}$ to $\Delta_{s;2}^{\rm res}$, then from $\Delta_{s;1}^{\rm off}$ to $\Delta_{s;1}^{\rm res}$. The excitation should out-couple with probabilities $|B|^2$ and $|A|^2$, realizing the output state $B|t_B\rangle + A|t_A\rangle$ where $|t_A\rangle, |t_B\rangle$ denote the temporally distinguishable basis states. Since the cavity decay ($\kappa_{wq}$) is a coherent out-coupling, the single photon is not a mixed state but a true superposition state.

We estimate the time scale for switching by calculating the coupling matrix element $J_\beta \equiv |\langle +_q|(\kappa_{sq}a_q^\dagger a_s)|\beta_s\rangle|$ ($\beta = u, v$) at resonances,
\begin{eqnarray}
	J_\beta = \kappa_{sq} \frac{g_s\Omega_s}{N_\beta} \cos\Theta,
\end{eqnarray}
and $N_\beta^2 = \mathcal{E}_{|\beta_s\rangle}^2\Omega_s^2/\hbar^2 + (\mathcal{E}_{|\beta_s\rangle}^2/\hbar^2 -\mathcal{E}_{|\beta_s\rangle}\Delta_s/\hbar - g_s^2]^2 + g_s^2\Omega_s^2$ and $\Theta$ is defined previously following Eq.~\ref{eq:J}. We impose the condition $J_u \approx J_v$ so that same sweep rate is used and the output pulse shapes at the two times are near identical for equal superposition states. Indeed, the parameters used in Fig.~\ref{fig:EigThreeLevel} are chosen for this purpose.

\begin{figure}[tb!]
\includegraphics[width=0.85\columnwidth,clip]{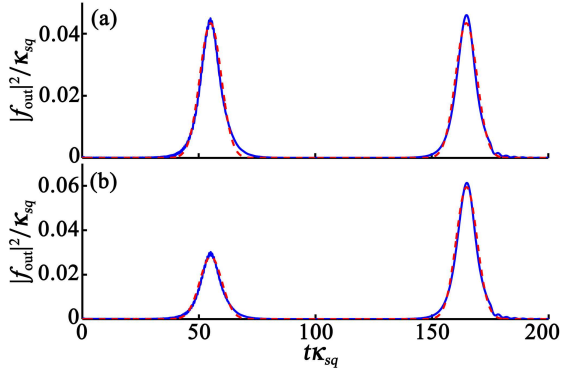}
\caption{(Color online) Single photon in two-time superposition state under no dissipation. Pulse shape of the output photon prepared with a linear sweep $\Delta_s(t)$ that varies from $\Delta_{s;2}^{\rm off} \rightarrow \Delta_{s;2}^{\rm res} \rightarrow \Delta_{s;1}^{\rm off} \rightarrow \Delta_{s;1}^{\rm res}$, for different initial state (a) $(|u_s\rangle+|v_s\rangle)/\sqrt{2}$, and (b) $(\sqrt{2}|u_s\rangle+|v_s\rangle)/\sqrt{3}$. The parameters follow Fig.~\ref{fig:EigThreeLevel} and $\kappa_{wq}/\kappa_{sq} = 4$. Dashed lines are Gaussian fits.}
\label{fig:DoublePulses}
\end{figure}

Applying the above prescription, we present the numerical simulations for initial equal superposition state in Fig.~\ref{fig:DoublePulses}(a), and unequal superposition $\sqrt{2}B = A = \sqrt{2/3}$ in (b). We should expect the integrated area of each pulse to be 0.5 in the first case, and 1/3 and 2/3 in the latter case. For $J_\alpha \sim 0.5$, a common linear sweep yields near-Gaussian superposition components with integrated areas of 0.502 and 0.495 for the equal superposition state, and 0.334 and 0.658 for the unequal state. The premature loss probability is less than $O(10^{-2})$. Following curve-fitting, the mode mismatch are $\xi = 0.012$ and $0.013$, respectively. Notably, the tuning time is much shorter than both the confinement time and the lapse between the pulses.

\begin{figure}[tb!]
\includegraphics[width=1\columnwidth,clip]{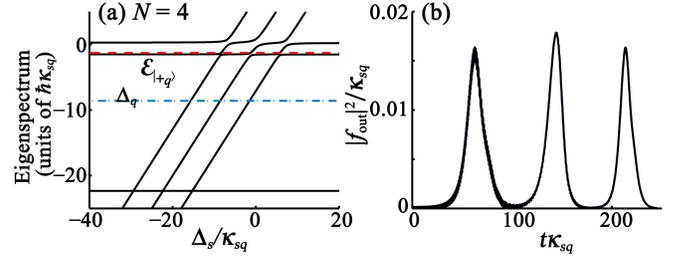}
\caption{(a) Eigensystem of the extended Hamiltonian $\mathcal{H}_{\rm sys}$ (Eq.~\ref{eq:HamS}) with $\Xi$-type, four-level atom $s$, for $\delta_q/\kappa_{sq} = -15, \Delta_q/\kappa_{sq} = -8.6, g_s/\kappa_{sq} = 1, \Omega_{s;i}/\kappa_{sq} = 5~(\forall i), g_q/\kappa_{sq} = 10$. Solid curves are the eigenenergies of the combined coupled-cavity system and the dashed line denote the energy of state $|+_q\rangle$. The dash-dotted line indicates the energy of atom $q$ ($= \hbar\Delta_q$), intersects at points (i.e. $\Delta_{s;i}^{\rm off}$) corresponding to two-photon resonance with atom $q$. The anticrossings along the dashed lines (i.e. $\Delta_{s;i}^{\rm res}$) correspond to resonances between eigenstates of system $s$ with $|+_q\rangle$. (b) Corresponding pulse shape of the output photon in three-time, unoptimized equal superposition state.}
\label{fig:SampleEig}
\end{figure}

\noindent \textit{Multi-time superposition}:-- We now generalize our results by considering a $\Xi$-type, $N$-level atom $s$ with states $|g_s\rangle, |e_{s;1}\rangle, |e_{s;2}\rangle, ..., |e_{s,N-1}\rangle$. The $|g_s\rangle \leftrightarrow |e_{s;1}\rangle$ transition is still coupled to cavity mode $s$ with rate $g_s$ whilst each excited-state pair (the $|e_{s;i}\rangle\leftrightarrow |e_{s;i+1}\rangle$ transition) is coupled via a quasi-resonant field with rate $\Omega_{s;i}$ and detuning $\delta_{s;i}$. The Hamiltonian at site $s$ becomes
\begin{eqnarray}
	\mathcal{H}_s/\hbar & = & \Delta_s|e_{s;1}\rangle\langle e_{s;1}| + \sum_{i=2}^{N-1}(\Delta_s + \delta_{s;i})|e_{s;i}\rangle\langle e_{s;i}| \label{eq:HamS}\\
	& + & g_s |e_{s;1}\rangle\langle g_s| a_s + \sum_{i=1}^{N-2} \Omega_{s;i} |e_{s;i+1}\rangle\langle e_{s;i}| + {\rm h.c.} \nonumber
\end{eqnarray}
Resonances between $N-1$ eigenstates of system $s$ and the one of the states $|\pm_q\rangle$ can be set up at different values of $\Delta_s = \Delta_{s;i}^{\rm res}$. In Fig.~\ref{fig:SampleEig}(a), we observe three anticrossings with state $|+_q\rangle$ for $N = 4$. Therefore, after initializing system $s$ in a superposition of these eigenstates, atom $s$ can be tuned to induce resonances in the coupled-cavity systems in series. Similarly, system $s$ becomes maximally decoupled from system $q$ via two-photon resonance with atom $q$. In principle, these mechanisms allow one to prepare arbitrary temporal superposition states with $N-1$ time-bins on demand. Three-time, unoptimized superposition state is shown in Fig.~\ref{fig:SampleEig}(b). To ensure that switching time is not too long, optimization is necessary to maximize the values of the coupling matrix element $J$. We emphasize that a huge parameter space $\{g_s,\Omega_{s;i},g_q,\delta_{s;i},\delta_q,\Delta_q\}$ should offer much room for this purpose.

For the sake of definitiveness, let us consider possible experimental realizations. Although we explicitly use $\Xi$-type atoms, one could use N-type $N=4$ and W-type $N=5$ atoms etc, in our discussion, other atomic configurations such as Y-type $N=4$, tripod $N=4$ and quadrupod $N=5$ can also be used. However, we will not delve further into discussing each of these schemes, apart from noting the abundance of systems for experimental work. In particular, the NV can realize $\Lambda$- and tripod-type schemes by using some or all available sublevels ($m = 0, \pm1$) in the spin triplet ($^3A$) ground state~\cite{manson06}. Although unsuitable for practical solid-state applications, rubidium~\cite{xiao95}, which has rich electronic ladder structures, and sodium with $Y$-type energy levels~\cite{gao00}, can be used for proof-of-concept experiments.

\section{Conclusion}
Advances in fabrication and novel cavity designs are nearing the stage where exploration and applications of multiple coupled atom-cavity systems are beginning to be accessible. We have studied the dynamic coupling between two evanescently-coupled cavities realized by tuning one of the intracavity atoms. Of practical interest to distributed QIP, this mediating atom-field system can be employed to realize high fidelity excitation confinement, $Q$-switching and reversible state transport of single photons. We also have shown that such control can be used to shape single photons and engineer near-Gaussian pulses that are distinct from the usual time-asymmetric photons emitted from a passive cavity-based system. 

The influence of cavity loss and atomic spontaneous emissions on the performance of $Q$ switching is quantitatively characterized, and we have shown that these schemes are compatible with the state-of-the-art cavity designs and atomic control capabilities in solid state. In particular, when implemented with diamond centres in PBG cavities, the effective cavity-$Q$ of qubit cavities can be switched in 10~ns time scale with $Q = 10^6$ with a success probability of 0.9. A shorter nanosecond switching is expected by using slot-waveguide cavities with $Q = 10^5$ for a probability of $> 0.99$. In the microwave regime, $1\mu$s switching is possible with stripline cavities of similar $Q$.

By applying $Q$-switching to an extended coupled-cavity system involving a multi-level atom, arbitrary temporally encoded superposition states can be prepared in a deterministic way. These integrated-photonic methods for generating high quality single photons and photonic qudits would be useful resources for numerous optical and general QIP applications. 

\section*{Acknowledgments}
We thank J. H. Cole, Z. W. E. Evans, S. J. Devitt, A. M. Stephens, S. Tomljenovic-Hanic, C. D. Hill, and M. I. Makin for helpful discussions. W.J.M. and K.N. acknowledge the support of QAP, HIP, MEXT, NICT and HP. C.H.S., A.D.G. and L.C.L.H. acknowledge the support of Quantum Communications Victoria, funded by the Victorian Science, Technology and Innovation (STI) initiative, the Australian Research Council (ARC), and the International Science Linkages program. A.D.G. and L.C.L.H. acknowledge the ARC for financial support (Projects No. DP0880466 and No. DP0770715, respectively).

\clearpage
\newpage
\onecolumngrid

\end{document}